\documentclass[10pt,journal,compsoc]{IEEEtran}
\usepackage{makecell}
\usepackage{array}
\newcolumntype{C}{>{\centering\arraybackslash}X}
\usepackage[nocompress]{cite}
\usepackage{graphicx}
\usepackage{epstopdf}
\usepackage{amsmath,amssymb,amsfonts}
\usepackage{epstopdf}
\usepackage{changes}
\usepackage{textcomp}
\usepackage{algorithmic}
\usepackage{xcolor}
\usepackage[ruled,lined,linesnumbered]{algorithm2e}
\usepackage{tabularx,booktabs}
\usepackage[flushleft]{threeparttable}

\usepackage{booktabs,ragged2e}
\usepackage{multirow}
\usepackage{subcaption}

\usepackage{bm}
\usepackage[colorlinks,
            linkcolor=blue,
            anchorcolor=black,
            citecolor=blue
            ]{hyperref}
\usepackage{amsthm}

\newtheorem{remark}{Remark}

\begin{document}
\renewcommand{\algorithmautorefname}{Algorithm}
\newcommand\MYhyperrefoptions{bookmarks=true,bookmarksnumbered=true,
pdfpagemode={UseOutlines},plainpages=false,pdfpagelabels=true,
colorlinks=true,linkcolor={black},citecolor={black},urlcolor={black},
pdftitle={Bare Demo of IEEEtran.cls for Computer Society Journals},
pdfsubject={Typesetting},
pdfauthor={Michael D. Shell},
pdfkeywords={Computer Society, IEEEtran, journal, LaTeX, paper,
             template}}
\title{Towards Secure Semantic Communications in the Presence of Intelligent Eavesdroppers}

\author{
Shunpu~Tang,
Yuhao~Chen, 
Qianqian~Yang,
Ruichen~Zhang,
Dusit~Niyato,
and Zhiguo Shi
\IEEEcompsocitemizethanks{
 \IEEEcompsocthanksitem This paper was in part presented at the IEEE Global Communications Conference (GLOBECOM), Cape Town, South Africa, 2024.\cite{10901326}. 
 \IEEEcompsocthanksitem S. Tang, Q. Yang and Z. Shi are with the College of Information Science and Electronic Engineering, Zhejiang University, Hangzhou, China (email: \{tangshunpu, qianqianyang20, shizg\}@zju.edu.cn). 
 \IEEEcompsocthanksitem Y. Chen is with the College of Control Science and Engineering (e-mail: csechenyh@zju.edu.cn)
 \IEEEcompsocthanksitem R. Zhang and D. Niyato are with the College of Computing and Data Science, Nanyang Technological University, Singapore (e-mail: \{ruichen.zhang, dniyato\}@ntu.edu.sg).}
}

\IEEEtitleabstractindextext{%
\begin{abstract}
\justifying
Semantic communication (SemCom) has emerged as a promising paradigm for enhancing communication efficiency in sixth-generation (6G) networks. However, the broadcast nature of wireless channels makes SemCom systems vulnerable to eavesdropping, which poses a serious threat to data privacy. Therefore, we investigate secure SemCom systems that preserve data privacy in the presence of eavesdroppers. Specifically, we first explore a scenario where eavesdroppers are intelligent and can exploit semantic information to reconstruct the transmitted data based on advanced artificial intelligence (AI) techniques. To counter this, we introduce novel eavesdropping attack strategies that utilize model inversion attacks and generative AI (GenAI) models. These strategies effectively reconstruct transmitted private data processed by the semantic encoder, operating in both glass-box and closed-box settings. Existing defense mechanisms against eavesdropping often cause significant distortions in the data reconstructed by eavesdroppers, potentially arousing their suspicion. To address this, we propose a semantic covert communication approach that leverages an invertible neural network (INN)-based signal steganography module. This module covertly embeds the channel input signal of a private sample into that of a non-sensitive host sample, thereby misleading eavesdroppers. Without access to this module, eavesdroppers can only extract host-related information and remain unaware of the hidden private content.  We conduct extensive simulations under various channel conditions in image transmission tasks. Numerical results show that while conventional eavesdropping strategies achieve a success rate of over 80\% in reconstructing private information, the proposed semantic covert communication effectively reduces the eavesdropping success rate to 0\%. Furthermore, results demonstrate that eavesdroppers can reconstruct host images with relatively good quality, achieving a peak signal-to-noise ratio (PSNR) of at most 27.50 dB, whereas the legitimate receiver can reconstruct the private image with the same perceptual quality as in scenarios without security mechanisms.

\end{abstract}
}
\maketitle

\IEEEdisplaynontitleabstractindextext
\IEEEpeerreviewmaketitle

\ifCLASSOPTIONcompsoc
\IEEEraisesectionheading{\section{Introduction}\label{sec:introduction}}
\else
\section{Introduction}
\label{Sec:intro}
\fi

\subsection{Background}

Semantic communication (SemCom) has emerged as a promising paradigm to support the upcoming sixth generation (6G) wireless networks\cite{Intellicise_Wireless, Semantic_survey}. Unlike traditional digital communication systems, which transmit raw data, SemCom systems aim to deliver the most relevant semantic information aligned with the intent of the receiver, which can reduce communication overhead and significantly improve the efficiency of massive information exchange\cite{semantic_IOT}. Such capabilities are particularly beneficial for realizing \textit{connected intelligence} \cite{chen2020connected} and \textit{artificial general intelligence (AGI)} \cite{AGI_6G} in 6G, playing a key role in facilitating machine-to-machine (M2M) and agent-to-agent communication \cite{zhang2025toward}.

Recent advances in SemCom have been driven by the rapid development of deep learning (DL) techniques. Specifically, the authors in \cite{DeepJSCC} first proposed a \textit{DL-based joint source-channel coding (DeepJSCC)} framework for image transmission, which extracts the semantic information of images and directly maps them into complex-valued channel input signals. Experimental results demonstrate that DeepJSCC effectively mitigates the cliff effect observed in traditional digital communication systems while significantly enhancing image reconstruction quality and transmission efficiency. Building on this work,  subsequent studies have further explored the potential of SemCom from various perspectives, such as the design of more powerful architectures for semantic encoders and decoders\cite{Attention_SemCom, SWINJSCC}, the integration of more advanced training strategies \cite{Shunpu_SemCom}, and achievement of more efficient resource allocation \cite{resource_SemCom,Energy_SemCom}. Specifically, the authors in \cite{Attention_SemCom} introduced channel-wise soft attention to focus on the most important semantic information under different channel conditions, thereby improving the adaptability, robustness, and versatility of the SemCom system. Similarly, the authors in \cite{SWINJSCC} addressed the limited performance of convolutional neural network (CNN) backbones by introducing a transformer-based architecture for the semantic encoder and decoder, which enhances the ability to capture long-range semantic dependencies and adapt to diverse channel conditions and transmission rates.

Moreover, the authors in \cite{Shunpu_SemCom} applied contrastive learning to train the SemCom system, ensuring that the transmitted semantic information remains distinguishable and robust against channel noises. Furthermore, the authors in \cite{resource_SemCom} investigated the QoE maximization problem in SemCom systems and proposed a joint framework to optimize the compression ratio, power allocation, and resource block allocation for multi-user scenarios. To improve the energy efficiency of SemCom systems, the authors in \cite{Energy_SemCom} explore a joint communication and computation strategy to minimize the energy consumption of the semantic encoder and decoder under computation, latency, and transmit power constraints.

Moreover, with the emergence of generative AI (GenAI) techniques, SemCom systems can be enhanced by leveraging the capabilities of GenAI models\cite{GenAI_SemCom}. In this direction, the authors in \cite{GAN_JSCC,tang2024evolving} explored the integration of generative adversarial networks (GANs) into SemCom systems and proposed GAN inversion-based semantic encoding to improve the fidelity of the received images. Additionally, diffusion models\cite{dhariwal2021diffusion} and large language models (LLM)\cite{brown2020language} have also been investigated to enable more efficient semantic information extraction and transmission \cite{chen2023commin, tang2024retrieval}. The authors in \cite{chen2023commin} proposed utilizing diffusion models as the semantic decoder, significantly enhancing the perceptual quality of the reconstructed images. The authors in \cite{tang2024retrieval} further combined multi-modal LLMs to extract semantic information from the source images and introduced a retrieval-augmented generation (RAG)  scheme at the diffusion-based semantic decoder to prevent hallucinations and improve the reconstruction quality.

\begin{figure}[!t]
    \centering
    \includegraphics[width=\linewidth]{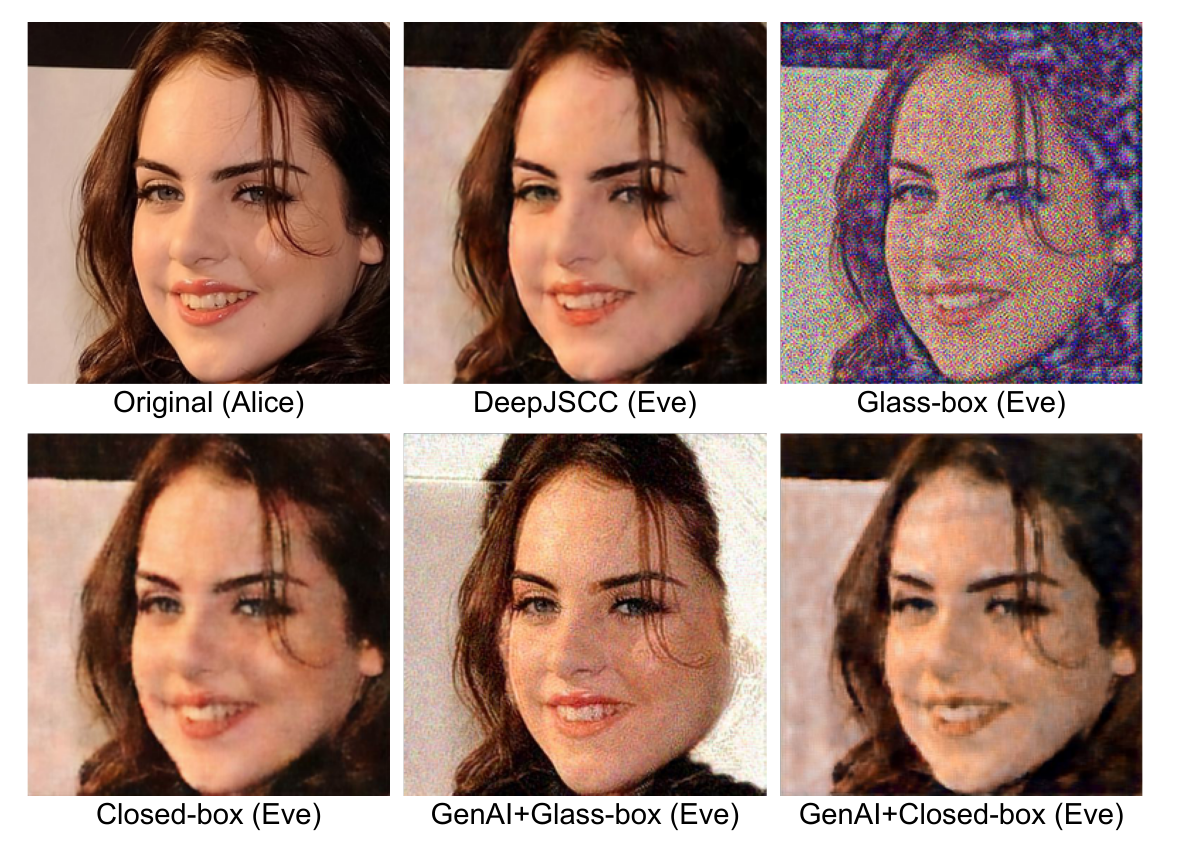}
    \caption{Visualization of the image reconstruction results by the proposed intelligent eavesdropping strategies under Rayleigh eavesdropping channels with an SNR of 5dB.}
    \label{fig:vis_eve}
\end{figure}

\textcolor{black}{However, such advancements in SemCom systems also bring new security challenges, making privacy leakage more likely than in conventional digital communication systems. This is because the semantic information typically exhibits lower redundancy and higher informational value and density. As a result, even partial semantic information leakage can lead to significant privacy risks\cite{Semantic_security_in_IOT,shen2023secure}. Moreover, in digital communication systems, various well-established physical layer security (PLS) techniques\cite{poor2017wireless} and digital encryption methods can be employed to safeguard the transmitted data, whereas, in SemCom systems, these methods are often difficult to implement effectively due to the end-to-end neural network-based joint source-channel coding design without quantization.  Therefore, even under poor channel conditions, an eavesdropper may still be able to decode and reconstruct the transmitted semantic information with the help of a powerful semantic decoder\cite{Semantic_security_zhaohui}, further increasing the risk of privacy leakage. Furthermore, even if the eavesdropper only has access to the semantic encoder or observes its input, it may still infer sensitive information by leveraging advanced learning-based techniques such as model inversion attacks (MIA) and GenAI techniques, where we refer to such eavesdroppers as \textit{intelligent eavesdroppers}\cite{10623738}. 
 As illustrated in \autoref{fig:vis_eve}, the original transmitted image can be easily reconstructed by eavesdroppers using the semantic decoder of DeepJSCC or the intelligent eavesdropping strategies proposed in this paper, including glass-box, closed-box MIA, and their GenAI-enhanced variants,  even under Rayleigh channels with SNR as low as 5 dB. These observations highlight the importance of investigating the security challenges in SemCom systems and developing effective defense mechanisms against eavesdroppers.}

\subsection{Related works}
To enhance SemCom security, researchers have explored various approaches, including encryption, PLS, and covert communication. For example, the authors in \cite{DeepJSCC_Encryption} integrated quantization operations and public-key encryption into SemCom, effectively protecting transmitted data against eavesdropping without significantly compromising system performance. Similarly, the authors in \cite{luo2023encrypted} designed an encryption module for semantic encoders/decoders with adversarial training to prevent eavesdroppers from reconstructing transmitted data without the encryption key. Extending this approach, homomorphic encryption was incorporated in the semantic encoder and decoder of a SemCom system to enable secure transmission and processing of encrypted data\cite{meng2025secure}. The authors in \cite{vis_protect} integrated a visual protection module before the semantic encoder, ensuring that eavesdroppers can only obtain distorted visual features that differ significantly from the original data without the corresponding deprotection module. The authors in \cite{weixuan_Differential_Privacy} applied a differential privacy (DP) mechanism to the semantic information before transmission, which uses two neural networks to approximate the non-invertible DP application and removal processes. 

Inspired by traditional PLS, the authors in \cite{Semantic_security_maojun, PLS_SemCom} introduced a novel secure mean square errors (MSE) loss function to ensure that the semantic decoder fails to decode the signal successfully when the channel conditions are worse than those encountered during training. This approach prevents eavesdroppers from exploiting the semantic decoder to overhear communications, as the channel conditions on the eavesdropping link are typically worse. However, this method may compromise the system’s robustness under poor channel conditions. In \cite{Secure_DeepJSCC_MultiEve}, the authors extended the secure SemCom system to a multi-eavesdropper scenario, where an improved loss function for SemCom training is proposed to enhance security against both colluding and non-colluding eavesdroppers. The authors in \cite{Semantic_security_weixaun} proposed a superposition coding scheme to maximize the symbol error probability (SER) of eavesdroppers while preserving the integrity of the main link, thereby ensuring approximately near-zero information leakage.

In addition to the above methods, the concept of covert communication\cite{covert_comm_2} has also been explored to enhance the security of SemCom systems by concealing the signals conveying private information from eavesdroppers. In this direction, the authors in \cite{du2024generative, liu2024learning} optimized the transmit power of both transmitted signals and friendly jamming signals to minimize eavesdroppers' detection probability. The authors in \cite{10570800} further proposed a novel covert semantic throughput (CST) metric to evaluate the covert communication capacity from the semantic perspective. Then they designed a full-duplex receiver to suppress the detection from eavesdroppers, and proposed a deep reinforcement learning to maximize the average CST improvement in low SNR scenarios.

\subsection{Motivation and Contributions}
\textcolor{black}{
Although the aforementioned and related works have made significant progress in enhancing the security of SemCom systems,  no comprehensive study has yet investigated both eavesdropping threats and corresponding defense mechanisms across diverse scenarios. Specifically, prior works primarily assume that eavesdroppers have access to the semantic decoder, while few studies systematically explore how eavesdroppers can infer private semantic information without decoder access, and also leverage advanced GenAI models to enhance their performance. Moreover, existing defense works using PLS and digital encryption often introduce detectable distortions in the reconstructed images by eavesdroppers\cite{vis_protect,Semantic_security_maojun,  weixuan_Differential_Privacy}, potentially raising suspicion from intelligent eavesdroppers and prompting them to perform active attacks such as jamming to disrupt communication\cite{pirayesh2022jamming}. While covert communication has been applied to SemCom\cite{covert_comm_2, du2024generative, liu2024learning}, these methods typically rely on deploying friendly jammers, which can degrade communication quality, making them unsuitable for scenarios that require continuous and reliable communication. This motivates us to extend the concept of covert communication to the semantic level, aiming to covertly transmit private semantic information while misleading eavesdroppers into believing they have successfully intercepted sensitive information.  
}

In this paper, we conduct a comprehensive study on the security of SemCom systems for image transmission under intelligent eavesdropping threats. We first propose MIA-based eavesdropping strategies to infer transmitted semantic information without requiring access to the semantic decoder. Specifically, we consider two scenarios: (i) \textit{glass-box} eavesdropping, where the eavesdropper has the knowledge of the semantic encoder's architecture and parameters; and (ii) \textit{closed-box} eavesdropping, where only the encoder’s input-output behavior can be observed\footnote{The terms glass-box and closed-box are used in place of the conventional white-box and black-box terminology throughout this paper.}. Motivated by recent advancements in GenAI models\cite{wang2024generative}, we employ GenAI techniques to further enhance the eavesdropping performance in both scenarios, enabling higher fidelity and more realistic reconstruction of private semantic content from intercepted signals.


\textcolor{black}{
 To mitigate the risks posed by such intelligent eavesdropping, we propose a novel semantic covert communication strategy developed upon a pre-trained SemCom, where a non-private image serves as a host to conceal private information. Instead of embedding private images directly at the image level, which incurs high computational overhead and is challenging for well-trained SemCom systems to process, we introduce an invertible neural network (INN)-based signal steganography module to embed the channel input signal of the private image into that of the host image before transmission. Since the channel input signal has a lower dimensionality than that of the original image, this approach reduces computational costs and enables efficient covert embedding. The output of the signal steganography module serves as the final channel input and is transmitted over noisy channels. On the legitimate receiver's side, the same signal steganography module is applied in reverse to extract the embedded private semantic information from the received signal, enabling accurate reconstruction of both the host and private images. On the other hand, eavesdroppers without access to this signal steganography module can reconstruct only the host image, thus ensuring the confidentiality of the private image.}

We conduct simulations on a facial image dataset to evaluate the proposed eavesdropping strategies and defense mechanisms under AWGN and Rayleigh channels. Specifically, we assess the performance of the glass-box and closed-box eavesdropping strategies, both with and without GenAI support, \textcolor{black}{by measuring the similarity between the eavesdropped and original transmitted images, as well as face recognition accuracy, which assesses whether the reconstructed private images can be identified as the same individuals as in the original images. We then evaluate the effectiveness of the proposed semantic covert communication against these eavesdropping strategies, demonstrating its ability to significantly reduce privacy leakage and enhance the security of SemCom systems in the presence of intelligent eavesdroppers.} In summary, the main contributions of this paper are as follows:
\begin{itemize}
    \item We propose novel intelligent eavesdropping strategies based on MIA and GenAI techniques to effectively reconstruct the transmitted images from intercepted signals under both glass-box and closed-box settings of semantic encoder access, posing significant privacy risks to SemCom systems.
    \item \textcolor{black}{To defend against the proposed intelligent eavesdropper, we introduce a semantic-level covert communication framework that utilizes an INN-based signal steganography module to embed the channel input of a private image into that of a non-sensitive host image. This ensures that eavesdroppers can only reconstruct host-related information while remaining unaware of the embedded private information.}
    \item  \textcolor{black}{We conduct extensive simulations to demonstrate that the proposed eavesdropping strategies achieve over 80\% success in reconstructing private content, while the semantic covert communication reduces the success rate to 0\%. Moreover,  the images reconstructed by eavesdroppers are highly similar to the host images with a PSNR up to 27.50dB, and the private images recovered by the legitimate receiver retain the same perceptual quality as in unsecured scenarios.}
\end{itemize}



\subsection{Organization}
The rest of the paper is organized as follows. \autoref{Sec:sys_model} presents the system model of SemCom and the eavesdropping model. \autoref{Sec:eve} introduces the proposed intelligent eavesdropping strategies based on MIA and GenAI models, followed by the proposed semantic covert communication in \autoref{Sec:covert}. \autoref{Sec:simulations} presents the simulation results to evaluate the proposed eavesdropping strategies and semantic covert communication. Finally, we conclude the paper and discuss future research directions in \autoref{Sec:conclusion}

\section{System Model}
\label{Sec:sys_model}
In this paper, we investigate a learning-based SemCom system for image transmission, involving two legitimate parties: a sender (i.e., Alice) and a receiver (i.e., Bob). Alice aims to transmit an image $\bm{x} \in \mathbb{R}^{N_\text{C} \times N_\text{H} \times N_\text{W}}$, where $N_\text{C}$, $N_\text{H}$, and $N_\text{W}$ represent the number of channels, height, and width of the RGB image, respectively. The source bandwidth of this image is given by $N = N_{\text{C}} \times N_\text{H} \times N_\text{W}$. Bob reconstructs the received image $\hat{\bm{x}}$.
\subsection{SemCom Model}
To reduce communication overhead, Alice employs a semantic encoder $\mathcal{E}: \mathbb{R}^N \to \mathbb{C}^k$ to extract key semantic information and directly maps it into a $k$-dimensional complex-valued channel input signal $\bm{z}$, which can be expressed as 
\begin{equation} 
\bm{z} = \mathcal{E}(\bm{x}; \theta_1), 
\end{equation} 
where $\theta_1$ represents the trainable parameters of the semantic encoder. The signal $\bm{z}$ is normalized to satisfy an average transmit power constraint $\bar{P}$\cite{DeepJSCC}, given by
\begin{equation} 
\frac{1}{k} \mathbb{E}[||\bm{z}||^2_2] \leq \bar{P}. 
\end{equation}
The bandwidth compression ratio (BCR) is defined as $k/N$. 

The channel signal $\bm{z}$ is then transmitted through a noisy channel, and the received signal at Bob is given by
\begin{equation} 
\hat{\bm{z}}_b = \bm{h}_b\odot \bm{z} + \bm{n}_b, 
\end{equation} 
where $\odot$ denotes the Hadamard product, $\bm{h}_b$ denotes the channel coefficient, and $\bm{n}_b$ represents the additive white Gaussian noise (AWGN) with zero mean and variance $\sigma^2_b$. Hence, the transmit signal-to-noise ratio (SNR) is given by $\text{SNR} = 10\log_{10}\bar{P} / \sigma^2_b$. Upon receiving $\hat{\bm{z}}_b$, Bob employs the semantic decoder $\mathcal{D}: \mathbb{C}^k \to \mathbb{R}^N$ to reconstruct the original image $\hat{\bm{x}}$ by mapping the received complex signal $\hat{\bm{z}}_b \in \mathbb{C}^k$ back to the real-valued image space $\hat{\bm{x}}_\text{b} \in \mathbb{R}^N$. The reconstruction process is given by:
\begin{equation} 
\hat{\bm{x}}_\text{b} = \mathcal{D}(\hat{\bm{z}}_b; \theta_2), 
\end{equation} 
where $\theta_2$ represents the trainable parameters of the semantic decoder.

\begin{figure*}[t]
    \centering
    \includegraphics[width=0.95\linewidth]{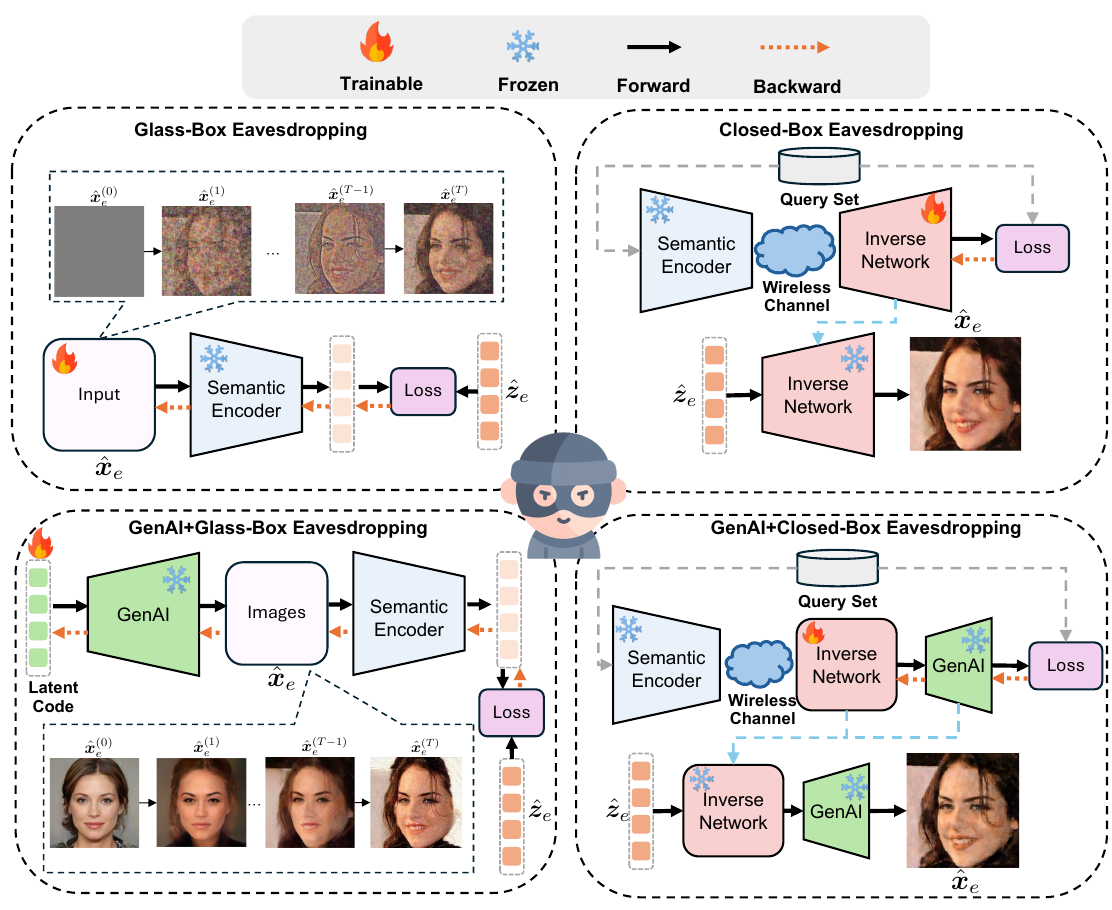}
    \caption{Visualization of the proposed intelligent eavesdropping including: 1) \textbf{Glass-box eavesdropping}, where Eve performs MIA to reconstruct the transmitted image using gradient descent. 2) \textbf{Closed-box eavesdropping}, where Eve trains an inverse network using its local dataset to perform MIA. 3) \textbf{GenAI+Glass-box eavesdropping}, where Eve introduces GenAI to produce high-quality candidates and inputs them into the semantic encoder to perform MIA.
     4) \textbf{GenAI+Closed-box eavesdropping}, where Eve trains an inverse network to generate latent codes for the GenAI model that can reconstruct the transmitted images with high fidelity.}
    \label{fig:overview_eve}
\end{figure*}
In the practical SemCom system, to jointly train the encoder $\mathcal{E}$ and decoder $\mathcal{D}$, we can minimize the expected risk over a training dataset $\mathcal{X}_{\text{train}} = \{\bm{x}^{(i)}\}_{i=1}^M$\cite{DeepJSCC,feifeiG_semantic}. The expected risk is defined as:
\begin{equation}
\mathcal{R}_M(\theta_1, \theta_2) = \mathbb{E}_{\bm{x} \sim \mathcal{X}_{\text{train}}} \left[ \mathcal{L}(\bm{x}, \hat{\bm{x}}_\text{b}) \right],
\end{equation}
where $\mathcal{L}(\bm{x}, \hat{\bm{x}}_\text{b})$ is the composite function of pixel-level and perceptual loss\cite{LPIPS}, given by
\begin{equation}
\begin{aligned}
    \mathcal{L}(\bm{x}, \hat{\bm{x}}_\text{b}) &= \underbrace{\lambda_1 \frac{1}{N} ||\bm{x} - \hat{\bm{x}}_\text{b}||_2^2}_{\text{Pixel-level loss}} \\ &+ 
\underbrace{\lambda_2 \sum_{l} \frac{1}{N_{H_l} N_{W_l}} \sum_{h=1,w=1}^{N_{H_l}, N_{W_l}} || \omega_l \odot (\bm{y}_{h,w}^l - \hat{\bm{y}}_{h,w}^l)||_2^2}_{\text{Perceptual loss}},
\end{aligned}
\end{equation}
where the perceptual loss evaluates high-level semantic similarity, ensuring that the reconstructed image aligns with the original image in terms of human visual perception. $\bm{y}_{h,w}^l$ and $\hat{\bm{y}}_{h,w}^l$  denote the feature vectors at spatial position $(h, w)$ in the $l$-th layer of a pre-trained feature extraction network, such as VGG16\cite{vgg}, and $\omega_l$ is a layer-wise weighting factor. $N_{H_l}$ and $N_{W_l}$ denote the height and width of the feature map at layer $l$, respectively. 

Finally, the optimization problem for training the semantic encoder and decoder is formulated as:
\begin{equation}
\{\theta_1^*, \theta_2^* \}= \arg\min_{\{\theta_1, \theta_2\}} \mathcal{R}_M(\theta_1, \theta_2),
\end{equation}
which can be optimized using gradient descent algorithms.

\subsection{Eavesdropping Model}
In wireless communication environments, the broadcast nature of wireless channels makes transmitted signals vulnerable to interception by unauthorized parties. Therefore, we consider a scenario where a passive eavesdropper (i.e. Eve), aims to intercept private information from the transmitted signal. The signal received by Eve can be expressed as
\begin{equation}
    \hat{\bm{z}}_e = \bm{h}_e \odot \bm{z} + \bm{n}_e,
\end{equation}
where $\bm{h}_e$ denotes the channel coefficient between Alice and Eve and $\bm{n}_e$ represents additive white Gaussian noise (AWGN) at Eve, following $\bm{n}_e \sim \mathcal{CN}(0, \sigma_e^2 \bm{I})$. Notably, we do not assume that the SNR of the eavesdropping link is significantly lower than that of Bob\cite{Semantic_security_maojun, 9053216}, thereby considering a more general and practical eavesdropping scenario. Next, Eve employs a potential reconstructed function, denoted as $\tilde{\mathcal{D}}(\cdot)$, to attempt to reconstruct the source image $\bm{x}$ by
\begin{equation}
    \hat{\bm{x}}_\text{e} = \tilde{\mathcal{D}}(\tilde{\bm{z}}; \tilde{\theta}),
\end{equation}
where $\tilde{\theta}$ represents the parameters of the reconstructed function. In this paper, we assume that Eve only has access to the semantic encoder. In the following, we will detail how it can be used to build the reconstructed function based on MIA and GenAI techniques and facilitate eavesdropping under various scenarios.
\begin{algorithm}[tbp]
    \caption{The proposed glass-box eavesdropping}  \label{Algo::wba}
    \KwIn{ Eavesdropping signal $\hat{\bm{z}}_e$, semantic encoder $\mathcal{E}(\cdot; \theta_1)$, channel coefficient $\bm{h}_e$, noise variance $\sigma_e^2$, learning rate $\eta$, maximum iterations $T_{\text{max}}$.}
    \KwOut{Reconstructed image $\hat{\bm{x}}_\text{e}$.}
    Initialize $\bm{x}^{(0)}_e \in \mathbb{R}^N$ randomly; \\
    \For{$t = 1$ \textbf{to} $T_{\text{max}}$}{
        Compute the forward function: $F(\bm{x}^{(t)}) = \bm{h}_e \cdot \mathcal{E}(\bm{x}^{(t)}; \theta_1)$; \\
        Compute the gradient of the loss: $\nabla \mathcal{L} = \frac{\partial}{\partial \bm{x}} \left( \frac{1}{2} \|\hat{\bm{z}}_e - F(\bm{x}^{(t)})\|_2^2 \right)$; \\
        Update the estimate: $\bm{x}^{(t+1)}_e = \bm{x}^{(t)}_e - \eta \nabla \mathcal{L}$; \\
        \If{$\|F(\bm{x}^{(t+1)}) - \hat{\bm{z}}_e\| < \epsilon$}{\textbf{break};}
    }
    Set $\hat{\bm{x}}_\text{e} = \bm{x}_e^{(t)}$; \\
    \Return $\hat{\bm{x}}_\text{e}$.
\end{algorithm}

\section{Intelligent Eavesdropping in SemCom}
\label{Sec:eve}
As illustrated in \autoref{fig:overview_eve}, we introduce intelligent eavesdropping strategies based on MIA and GenAI techniques. Specifically, we consider two scenarios of the glass-box and the closed-box eavesdropping, which depend on the level of knowledge Eve has about the semantic encoder.

\subsection{Glass-box Eavesdropping}
In this scenario, Eve has full access to the architecture of the semantic encoder and its parameters. This situation can arise when the SemCom system is publicly available, such as through open-source distribution or collaborative learning frameworks like \textit{federated learning (FL)} \cite{Semantic_FL, Semantic_FL_2}, where the semantic encoder and its parameters are accessible to participants.

Given full knowledge of the semantic encoder, Eve can build the decoding function by using the MIA technique, which aims to reconstruct the source image $\bm{x}$ from the intercepted signals $\hat{\bm{z}}_e$ by solving the inverse problem that maximizes the posterior probability $p(\bm{x}|\hat{\bm{z}}_e)$, given by
\begin{equation}
\label{eq:map}
\begin{aligned}
    \hat{\bm{x}}_\text{e} &=\tilde{\mathcal{D}}(\bm{z};\theta_1)= \arg\max_{\bm{x}} p(\bm{x}|\hat{\bm{z}}_e) \\
    &\propto \arg\max_{\bm{x}} p(\hat{\bm{z}}_e|\bm{x}) p(\bm{x}),
\end{aligned}
\end{equation}
where $p(\hat{\bm{z}}_e|\bm{x})$ represents the likelihood of observing $\hat{\bm{z}}_e$ given $\bm{x}$, and $p(\bm{x})$ is the prior distribution of the source $\bm{x}$. Given AWGN noises, the likelihood term $p(\hat{\bm{z}}_e|\bm{x})$ can be written as 
\begin{equation}
\label{eq:p_ze_x}
    p(\hat{\bm{z}}_e|\bm{x}) \propto \exp\left(-\frac{\|\hat{\bm{z}}_e - F(\bm{x})\|_2^2}{2\sigma_e^2}\right),
\end{equation}
where we define $F(\bm{x})=\bm{h}\odot \mathcal{E}(\bm{x};\theta_1)=\bm{h}\odot\bm{z}$ as a \textit{forward function}.
It is noted that if the prior $p(\bm{x})$ is unknown, the optimization problem reduces to maximize the likelihood, which is equivalent to minimizing the MSE between $\hat{\bm{z}}_e$ and $F(\bm{x})$, which can be expressed as
\begin{equation}
    \hat{\bm{x}}_\text{e} = \arg\min_{\bm{x}\in \mathbb{R}^{N}} \|\hat{\bm{z}}_e - F(\bm{x})\|_2^2.
\end{equation}
Hence, we can perform the gradient descent and iteratively update $\bm{x}$ to solve this problem. The key steps are summarized in \autoref{Algo::wba}.

\subsection{Closed-box Eavesdropping}

In the closed-box scenario, Eve does not have direct knowledge of the internal architecture or parameters of the semantic encoder, and cannot obtain the explicit form of the likelihood function $p(\hat{\bm{z}}_e|\bm{x})$, which prevents Eve from optimizing the posterior probability $p(\bm{x}|\tilde{\bm{z}}_e)$ by performing gradient-based optimization as in the case of glass-box eavesdropping. Instead, Eve can still observe the input-output behavior of the semantic encoder via the application programming interface (API) provided by the system owner\cite{papernot2017practical}, and employ a data-driven strategy to \emph{approximate} the posterior distribution $p(\bm{x}|\tilde{\bm{z}}_e)$ in an implicit manner. 
\begin{algorithm}[tbp]
    \caption{The proposed closed-box eavesdropping}
    \label{Algo::bba}
    \KwIn{
        closed-box encoder API $\mathcal{E}(\cdot)$,
        query set $\{\bm{x}_i\}_{i=1}^M$,
        inverse network $\tilde{\mathcal{D}}(\cdot;\phi)$,
        learning rate $\eta$,
        maximum training epochs $T_{\text{max}}$
    }
    \KwOut{Reconstructed image(s) $\hat{\bm{x}}_\text{e}$}

    \textbf{Stage 1: Data Collection} \\
    Construct dataset $\mathcal{X}_e \leftarrow \{\}$; \\
    \For{$i = 1$ \textbf{to} $M$}{
        Query the closed-box encoder: $\bm{z}_i = \mathcal{E}(\bm{x}_i;\theta_1)$; \\
        Apply simulated channel: $\tilde{\bm{z}}_i = \bm{h}\odot\bm{z}_i+\bm{n}$; \\
        Record pair $\bigl(\bm{x}_i, \tilde{\bm{z}}_i\bigr)$ in $\mathcal{X}_e$;
    }

    \BlankLine
    \textbf{Stage 2: Inverse Network Training} \\
    Initialize parameters $\phi$ randomly; \\
    \For{$t = 1$ \textbf{to} $T_{\text{max}}$}{
        \For{\textbf{each} minibatch $\{(\bm{x}_j, \tilde{\bm{z}}_j)\}$ from $\mathcal{X}_e$}{
            Compute inverse outputs: $\hat{\bm{x}}_j^{(t)} = \tilde{\mathcal{D}}(\tilde{\bm{z}}_j; \phi)$; \\
            Calculate the MSE loss: $\mathcal{L}(\phi) = \frac{1}{\lvert\mathrm{minibatch}\rvert} \sum_{j} \bigl\|\hat{\bm{x}}_j^{(t)} - \bm{x}_j\bigr\|_2^2;$ \\
            Compute the gradient $\nabla_{\phi}\mathcal{L}(\phi)$ and update: $\phi \leftarrow \phi - \eta \,\nabla_{\phi}\mathcal{L}(\phi);$
        }
    }
    Set $\phi^* \leftarrow \phi$; \\

    \BlankLine
    \textbf{Stage 3: Eavesdropping} \\
    \For{\textbf{each} intercepted signal $\tilde{\bm{z}}_e$}{
        Reconstruct: $\hat{\bm{x}}_\text{e} = \tilde{\mathcal{D}}\bigl(\tilde{\bm{z}}_e; \phi^*\bigr);$
    }

    \Return $\hat{\bm{x}}_\text{e}$
\end{algorithm}

Specifically, Eve first queries the closed-box semantic encoder with a collection of known $M$ samples $\{\bm{x}_i\}^{M}_{i=1}$ and records the corresponding outputs $\tilde{\bm{z}}_i$ after passing them through a simulated channel. In this way, Eve constructs a dataset
\begin{equation}
    \mathcal{X}_e = \{ (\bm{x}_i, \tilde{\bm{z}}_i) \}_{i=1}^M
\end{equation}
which captures the input-output behavior of the semantic encoder.
With this collected dataset, Eve trains a parametric inverse network $\tilde{\mathcal{D}}(\cdot; \phi)$, parameterized by $\phi$, to approximate the inverse mapping
\begin{equation}
\label{eq:posterior-black}
    \hat{\bm{x}}_\text{e} = \tilde{\mathcal{D}}(\tilde{\bm{z}}_e; \phi)= \arg\max_{\bm{x}} p(\bm{x} \mid \tilde{\bm{z}}_e).
\end{equation}
Therefore, the training objective for the inverse network can be formulated as:
\begin{equation}
\label{eq:blakc_object}
\phi^* = \arg\min_{\phi} \mathbb{E}_{(\bm{x}, \tilde{\bm{z}}_e) \sim \mathcal{X}_e} \left[ \| \tilde{\mathcal{D}}(\tilde{\bm{z}}_e; \phi) - \bm{x} \|_2^2 \right].
\end{equation}
Once the training is completed, Eve can perform eavesdropping by simply feeding any newly intercepted code $\tilde{\bm{z}}_e$ in the $\mathcal{D(\cdot,\phi^*)}$. We summarize the whole process of the proposed closed-box eavesdropping in \autoref{Algo::bba}.

\begin{remark}
 From a Bayesian perspective, both the glass-box and closed-box attacks aim to solve the same posterior maximization problem in~\eqref{eq:map}. However, in the glass-box setting, Eve can explicitly characterize $p(\tilde{\bm{z}}_e|\bm{x})$ due to her direct access to the forward function $F(\bm{x})$, and hence can employ gradient-based inversion. In contrast, in the closed-box setting, Eve cannot derive the closed-form expression of $F(\bm{x})$ and must therefore implicitly learn the inverse mapping using a parametric model, as formulated in~\eqref{eq:blakc_object}. 

\end{remark}

\subsection{Enhanced Eavesdropping with GenAI Models}
Motivated by the success of GenAI models in approximating complex data distributions, we further explore the use of GenAI models to provide an estimated prior in~\eqref{eq:map} of the original data to facilitate eavesdropping. Specifically, for a well-trained GAN, the generative process is
\begin{equation}
    \bm{x} = G(\bm{s}), \quad \bm{s} \sim p(\bm{s}) 
\end{equation}
where $\bm{s}$ is the latent code sampled from a prior distribution $p_s(\bm{s})$. Next, we discuss how GenAI models can be integrated into the eavesdropping process under both glass-box and closed-box settings. 
\subsubsection{GenAI for Glass-box Eavesdropping}
In the case of eavesdropping via the glass-box attack, we can write the inverse problem as
\begin{equation}
\hat{\bm{s}}=\arg\max_{\bm{s}} p \bigg(\bm{z_e}|G(\bm{s})\bigg)p(\bm{s}).
\end{equation}
Similar to ~\eqref{eq:p_ze_x}, we have 
\begin{equation}
\begin{aligned}
        p(\bm{z}|\bm{s}) & \propto \exp \bigg( \frac{-||\hat{\bm{z}}_e-F\big(G(\bm{s})\big)||^2_2}{2\sigma_e^2}\bigg), \\
\end{aligned}
\end{equation}
Since the latent variable $\bm{s}$ is often modeled as following a Gaussian prior, i.e., $\bm{s} \sim \mathcal{N}(\bm{0}, \bm{I})$,  we then have
\begin{equation}
    p(s) \propto \exp \bigg( -\frac{1}{2} ||\bm{s}||^2_2 \bigg).
\end{equation}
Thus, substituting the likelihood and prior into the MAP formulation, the optimal latent variable estimate can be derived by optimizing the following objective function
\begin{equation}
    \bm{s}^* = \arg\min_{\bm{s}} \bigg ( \frac{1}{2\sigma_e^2} \big\|\hat{\bm{z}}_e - F(G(\bm{s}))\big\|_2^2 + \frac{1}{2} \big\|\bm{s}\big\|_2^2 \bigg).
\end{equation}
 Once the optimal latent code $\bm{s}^*$ is found, the reconstructed version is given by
\begin{equation}
\hat{\bm{x}}_\text{e} = G(\hat{\bm{s}}).
\end{equation}
\subsubsection{GenAI for Closed-box Eavesdropping}
In the case of closed-box eavesdropping, a well-trained GenAI model, such as StyleGAN, can also be used to further enhance the eavesdropping performance.  Specifically,  when we adopt GANs as the GenAI model, instead of training an inverse network $\tilde{\mathcal{D}}(\cdot; \phi)$ to approximate the inverse mapping and output the reconstructed image,  we modify its architecture so that the output consists of two components: a latent code $\bm{s} \in \mathbb{R}^{d_s}$ and a noise vector $\bm{n} \in \mathbb{R}^{d_n}$, i.e.,
\begin{equation}
(\bm{s}, \bm{n}) = \tilde{\mathcal{D}}(\tilde{\bm{z}}_e; \phi),
\end{equation}
where $\bm{s}$ matches the latent space of StyleGAN, and $\bm{n}$ is used to replace or control the per-layer noise typically injected within the StyleGAN generator. In this design, we adapt the generator $G(\cdot, \cdot)$ to accept both the latent code and the noise vector, generating the reconstructed image as:
\[
\hat{\bm{x}}_\text{e} = G(\bm{s}, \bm{n}).
\]
This allows Eve to control not only the semantic content via $\bm{s}$ but also the stochastic texture details via $\bm{n}$. As such, it can improve the fidelity and consistency of reconstructed images, especially for high-frequency visual information.

Accordingly, the inverse network can be trained by minimizing the reconstruction error between the generated and ground-truth images while keeping the StyleGAN generator frozen. The training objective is:
\begin{equation}
    \phi^* = \arg\min_{\phi} \mathbb{E}_{(\bm{x}, \tilde{\bm{z}}_e) \sim \mathcal{X}_e} \left[ \left\| G(\tilde{\mathcal{D}}(\tilde{\bm{z}}_e; \phi)) - \bm{x} \right\|_2^2 \right].
\end{equation}
After training, given any intercepted signal $\tilde{\bm{z}}_e$, Eve can obtain $(\bm{s}, \bm{n}) = \tilde{\mathcal{D}}(\tilde{\bm{z}}_e; \phi^*)$ and reconstruct the image as $\hat{\bm{x}}_\text{e} = G(\bm{s}, \bm{n})$.

\begin{figure*}[!ht]
  \centering
     \includegraphics[width=0.98\linewidth]{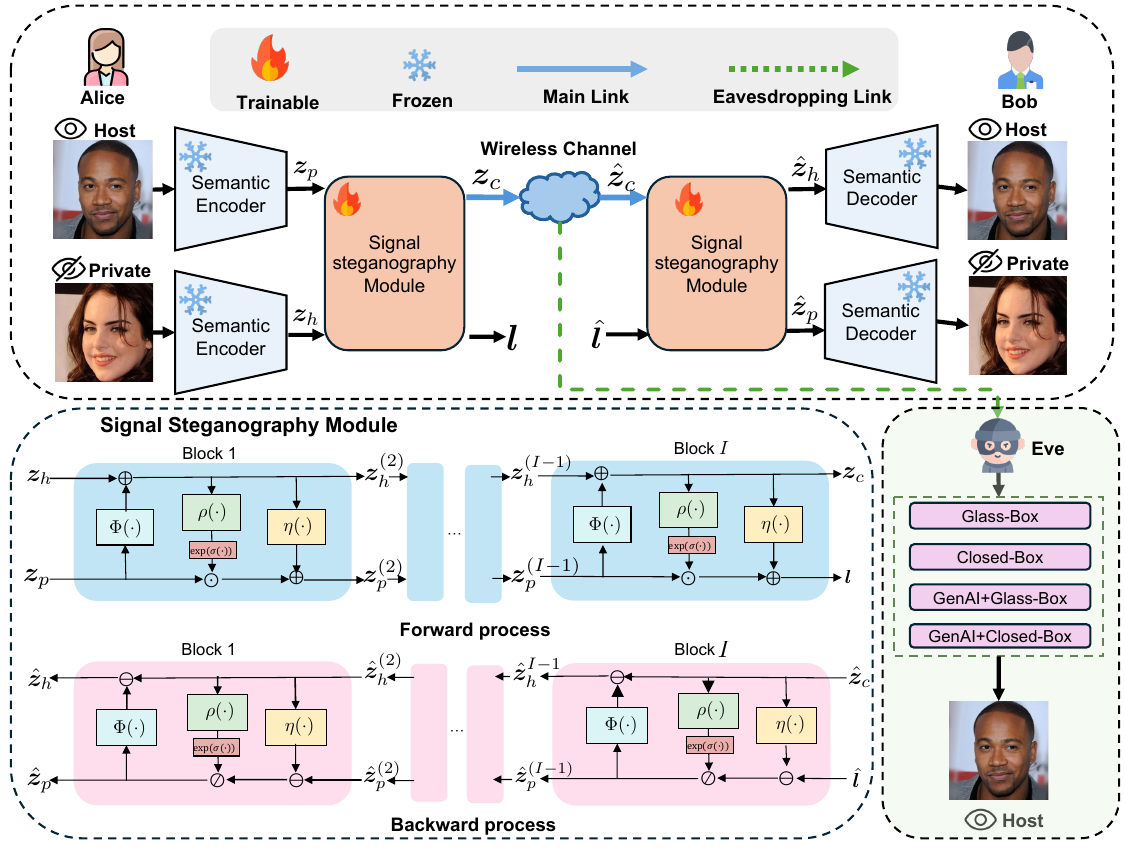}
    \caption{Illustration of the proposed semantic covert communication.}
    \label{fig:INN}
\end{figure*}

\section{Semantic Covert Communication}
\label{Sec:covert}
In this section, we introduce the proposed semantic covert communication, which employs signal steganography to embed private semantic information within a non-sensitive host image to deceive Eve. Specifically, we first present the whole pipeline of semantic covert communication with signal steganography, followed by a detailed description of the design of the signal steganography module, as well as the training procedure and loss function.

\subsection{Overview of Semantic Covert Communication}
As shown in \autoref{fig:INN}, an INN-based signal steganography module is deployed at both Alice's and Bob's sides to conceal the channel input signal corresponding to the private image within that of the host image. To simplify the presentation, we use $\bm{x}_h$ and $\bm{x}_p$ to denote the host image and private image, respectively. These two images are first input into the semantic encoder and obtain the corresponding channel input signals $\bm{z}_h$ and $\bm{z}_p$, given by
\begin{equation}
    \begin{aligned}
        \bm{z}_h = \mathcal{E}(\bm{x}_h; \theta_1), \quad
         \quad \bm{z}_p = \mathcal{E}(\bm{x}_p; \theta_1).
    \end{aligned}
\end{equation}
Then, the signal steganography module is applied to these channel input signals, combining them into a single container signal $\bm{z}_{c}$,  given by
\begin{equation} 
(\bm{z}_{c}, \bm{l}) = \mathcal{S}(\bm{z}_h,\bm{z}_p; \psi),
\end{equation}
where $\mathcal{S}(\cdot)$ represents the INN-based steganography module with parameter $\phi$, and $\bm{l}$ denotes the information lost during this steganography process. It is important to note that  $\bm{z}_c$ has the same dimension as both $\bm{z}_p$ and $\bm{z}_h$, which means that the BCR remains unchanged in this process.

Following this, the container signal $\bm{z}_c$ is transmitted over a noisy communication channel. Next, we use $\hat{\bm{z}}_c$ and $\hat{\bm{z}}_{c,e}$ to denote the received signal at Bob and Eve, respectively. To recover the original channel input signals, Bob applies the inverse operation of the same signal steganography module, given by
\begin{equation}
        (\hat{\bm{z}}_h, \hat{\bm{z}}_p)= \mathcal{S}^{-1}(\hat{\bm{z}}_c, \hat{\bm{l}}; \psi),
\end{equation}
where $\hat{\bm{l}}$ is the estimated lost information at Bob, which is either a predefined constant value or sampled from a given distribution. Subsequently, Bob can reconstruct both the host and private images, $\hat{\bm{x}}_h$ and $\hat{\bm{x}}_p$, respectively, using the semantic decoder. In contrast, Eve does not have the same signal steganography module and can obtain only information about $\bm{x}_h$ by employing the aforementioned eavesdropping strategies.
\begin{algorithm}[tbp]
    \caption{Training Procedure of INN-based Signal Steganography}  \label{algo:INN}
    \KwIn{Training dataset $\mathcal{X}_{\text{INN}}$, encoder $\mathcal{E}$, decoder $\mathcal{D}$, steganography module $\mathcal{S}$, learning rate $\eta$, batch size $B$, epochs $T$}
    \KwOut{Optimized parameters $\psi^*$}
    \For{$t = 1$ to $T$}{
        \For{each mini-batch $\{(\bm{x}_h, \bm{x}_p)\}_{i=1}^{B} \sim \mathcal{X}_{\text{INN}}$}{
            Compute channel inputs: $\bm{z}_h = \mathcal{E}(\bm{x}_h; \theta_1)$, $\bm{z}_p = \mathcal{E}(\bm{x}_p; \theta_1)$\;
            Generate container signal: $(\bm{z}_c, \bm{l}) = \mathcal{S}(\bm{z}_h, \bm{z}_p; \psi)$\;
            Transmit $\bm{z}_c$ over a noisy channel\;
            Recover signals using inverse steganography: $(\hat{\bm{z}}_h, \hat{\bm{z}}_p) = \mathcal{S}^{-1}(\hat{\bm{z}}_c, \hat{\bm{l}}; \psi)$, where $\hat{\bm{l}}$ is an estimate of $\bm{l}$\;
            Decode images: $\hat{\bm{x}}_h = \mathcal{D}(\hat{\bm{z}}_h; \theta_2)$, $\hat{\bm{x}}_p = \mathcal{D}(\hat{\bm{z}}_p; \theta_2)$\;
            Compute total loss: $\mathcal{L}_{\text{total}} = \mathcal{L}_{\text{forward}} + \mathcal{L}_{\text{backward}} + \mathcal{L}_{\text{privacy}}$\;
            Update parameters: $\psi \leftarrow \psi - \eta \nabla_{\psi} \mathcal{L}_{\text{total}}$\;  
        }
    }
    \Return $\psi^*$
\end{algorithm}

\subsection{Design of Signal Steganography Module}
To obtain the channel input signal $\bm{z}_c$ while ensuring the accurate recovery of the original signals, we adopt an INN-based architecture for the signal steganography module. This is because INN enables reversible operations, ensuring a precise reconstruction of the original signals from the INN output signals.


As shown in \autoref{fig:INN}, the proposed INN-based signal steganography module consists of multiple invertible blocks. Following the design of invertible blocks in \cite{ISN}, we use additive affine transformations to extract the features. Mathematically, let $ \bm{z}_h^i $ and $ \bm{z}_p^i $ be the input signals at the $ i $-th block, and $ \bm{z}_h^{(i+1)} $ and $ \bm{z}_p^{(i+1)} $ be the corresponding outputs. The forward operation of the $ i $-th invertible block is given by:
\begin{equation}
\label{eq:inn_forward}
    \begin{aligned}
        \bm{z}_h^{(i+1)} &= \bm{z}_{h}^{(i)} + \Phi(\bm{z}_{p}^{(i)}), \\
        \bm{z}_p^{(i+1)} &= \bm{z}_{p}^{(i)} \odot \exp\bigg(\rho(\bm{z}_h^{(i+1)})\bigg) + \eta(\bm{z}_h^{(i+1)}),
    \end{aligned}
\end{equation}
where $ \Phi(\cdot) $, $ \rho(\cdot) $, and $ \eta(\cdot) $ are additive affine transformations implemented using convolutional blocks. 

Next, we detail the forward operation $\mathcal{S}(\bm{z}_h,\bm{z}_p;\psi)$. Starts from the initial input signal $ \bm{z}_h^{(1)} = \bm{z}_h $ and $ \bm{z}_p^{(1)} = \bm{z}_p $, and iteratively applies the invertible transformations for $ I $ blocks, yielding the final container signal and lost information:
\begin{equation}
    \bm{z}_c = \bm{z}_h^{(I)}, \quad \bm{l} = \bm{z}_p^{(I)}.
\end{equation}

Accordingly, the backward operation of the INN-based signal steganography module denoted as $ \mathcal{S}^{-1}(\bm{z}_c, \hat{\bm{l}}; \psi) $, reconstructs the original channel input signals $ \hat{\bm{z}}_h $ and $ \hat{\bm{z}}_p $ from the received container signal $ \hat{\bm{z}}_c $ and the estimated lost information $ \hat{\bm{l}} $. The inverse operation mirrors the forward transformation but is applied in the reverse order of~\eqref{eq:inn_forward}. Specifically,  we first initialize $\hat{\bm{z}}_h^{(I)}=\hat{\bm{z}}_c$ and $\hat{\bm{z}}_p^{(I)} = \hat{\bm{l}}$. Then, the backward operation iteratively applies the following transformations for each invertible block:
\begin{equation}
\label{eq:INN_backward}
    \begin{aligned}
        \hat{\bm{z}}_p^{(i)} &= \bigg (\hat{\bm{z}}_p^{(i+1)} - \eta(\hat{\bm{z}}_h^{(i+1)})\bigg) \odot \exp\bigg(-\rho(\hat{\bm{z}}_h^{(i+1)})\bigg), \\
        \hat{\bm{z}}_h^{(i)} &= \hat{\bm{z}}_h^{(i+1)} - \Phi(\hat{\bm{z}}_p^{(i)}).
    \end{aligned}
\end{equation}
By successively applying~\eqref{eq:INN_backward} across all $ I $ invertible blocks, the original channel input signals can be derived.


\subsection{Training Procedure and Loss Function}
To mitigate the risk of privacy leakage, we design a training procedure and loss function for the proposed signal steganography module. Firstly, in the forward process of the signal steganography module, we aim to minimize the discrepancy between the container signal $\bm{z}_c$ and the channel input signal $\bm{z}_h$ of the non-sensitive host image, ensuring that Eve can only decode to the information related to $\bm{x}_h$. Moreover, we note that $\bm{l}$ is not transmitted to Bob. Instead, Bob uses either a predefined constant value or a sample from a given distribution to approximate $\bm{l}$. Therefore, it is also necessary to minimize the difference between $\bm{l}$ and $\hat{\bm{l}}$ to ensure Bob can successfully reconstruct the original channel input signals $\bm{z}_h$ and $\bm{z}_p$ through the backward process of the INN. For these reasons, the loss function of the forward process of INN can be expressed as
\begin{equation}
    \mathcal{L}_{\text{forward}} = \lambda_1||\bm{z}_c-\bm{z}_h||^2_2+\lambda_2||\bm{l}-\hat{\bm{l}}||^2_2,
\end{equation}
where $\lambda_1>0$ and $\lambda_2>0$ are hyperparameters to balance these two terms.

At Bob's end, the objective is to reconstruct $\bm{z}_p$ and $\bm{z}_h$ using the backward process of the signal steganography module. To achieve this, we define the loss function for the backward process as follows:
\begin{equation} \mathcal{L}_{\text{backward}} = \lambda_3||\bm{z}_p - \hat{\bm{z}}_p||^2_2 + \lambda_4||\bm{z}_h - \hat{\bm{z}}_h||^2_2, \end{equation}
where $\lambda_3 > 0$ and $\lambda_4 > 0$ are hyperparameters that control the trade-off between the reconstruction accuracy of $\bm{z}_p$ and $\bm{z}_h$. 

\begin{figure*}[!t]
    \centering
    \begin{subfigure}[t]{0.30\textwidth}
        \centering
        \includegraphics[width=\textwidth]{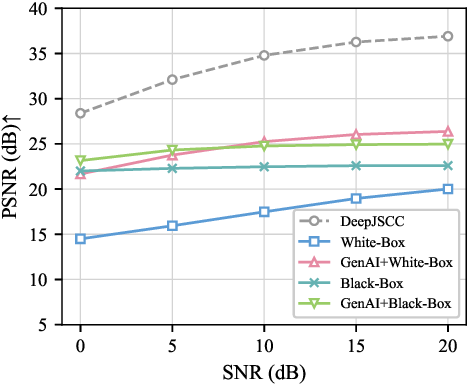}
        \caption{PSNR}
        \label{fig:sub1}
    \end{subfigure}
    \hfill
    \begin{subfigure}[t]{0.30\textwidth}
        \centering
        \includegraphics[width=\textwidth]{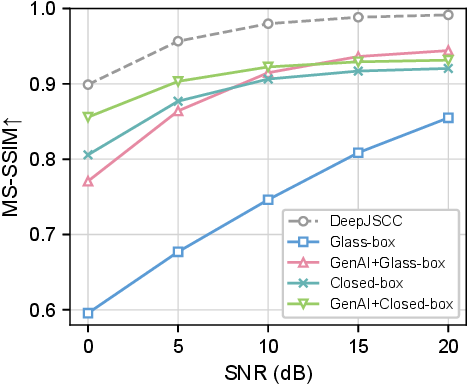}
        \caption{MS-SSIM}
        \label{fig:sub2}
    \end{subfigure}
        \hfill
    \begin{subfigure}[t]{0.294\textwidth}
        \centering
        \includegraphics[width=\textwidth]{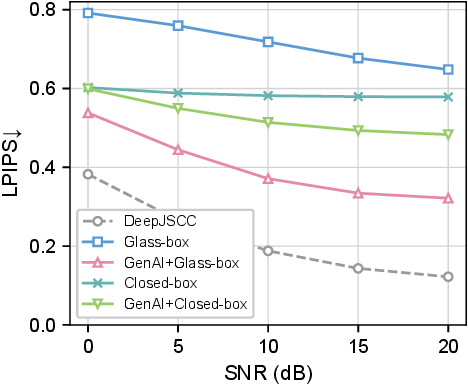}
        \caption{LPIPS}
        \label{fig:sub3}
    \end{subfigure}
    \caption{Performance of the proposed eavesdropping strategies under AWGN channels, where SNR varies from 0 dB to 20 dB.}
    \label{fig:AWGN_Eve}
\end{figure*}
\begin{figure*}[!t]
    \centering
    \begin{subfigure}[t]{0.30\textwidth}
        \centering
        \includegraphics[width=\textwidth]{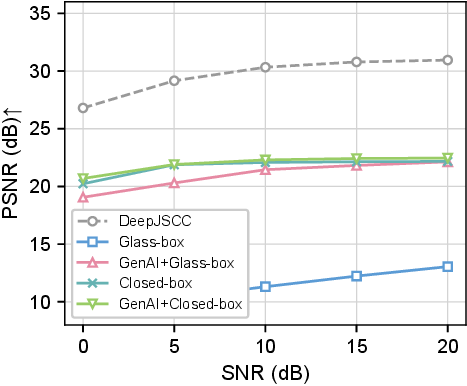}
        \caption{PSNR}
        \label{fig:sub1}
    \end{subfigure}
    \hfill
    \begin{subfigure}[t]{0.30\textwidth}
        \centering
        \includegraphics[width=\textwidth]{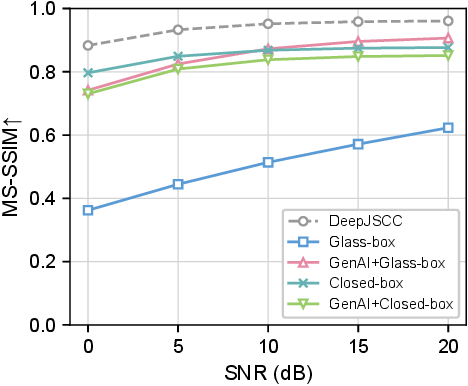}
        \caption{MS-SSIM}
        \label{fig:sub2}
    \end{subfigure}
        \hfill
    \begin{subfigure}[t]{0.294\textwidth}
        \centering
        \includegraphics[width=\textwidth]{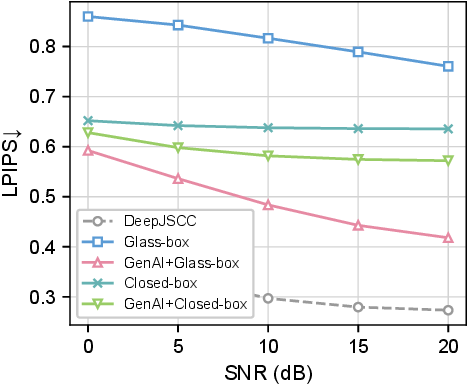}
        \caption{LPIPS}
        \label{fig:sub3}
    \end{subfigure}
    \caption{Performance of the proposed eavesdropping strategies under Rayleigh channels, where SNR varies from 0 dB to 20 dB. }
    \label{fig:Rayleigh_Eve}
\end{figure*}

\begin{figure}[!t]
    \centering
    \includegraphics[width=0.9\linewidth]{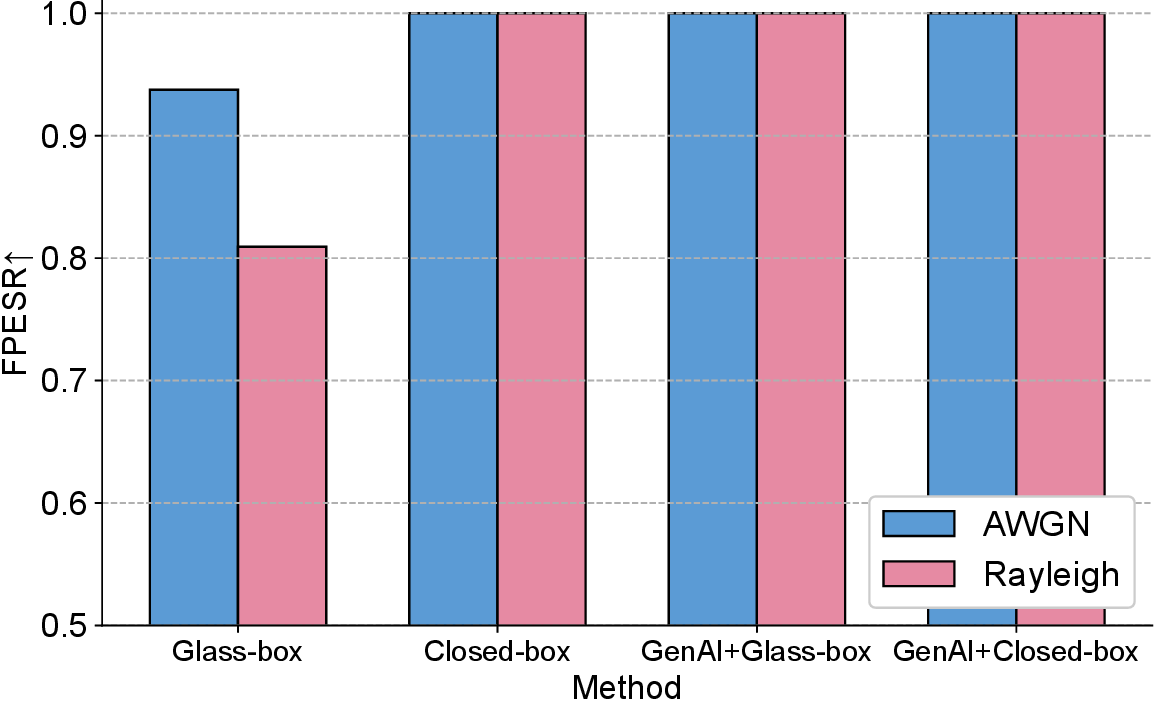}
    \caption{FPPSER of the proposed eavesdropping strategies under different channels with SNR of 5dB.}
    \label{fig:fpesr}
\end{figure}
Moreover, to ensure the reconstruction quality of the private image at Bob's end, we introduce an additional loss term, referred to as privacy consistency loss. This term enforces consistency between the original private image $\bm{x}_p$ and its reconstructed version $\hat{\bm{x}}_p$, thereby mitigating distortions caused by channel noise and transformation processes. The privacy consistency loss is formulated as
\begin{equation} \mathcal{L}_{\text{privacy}} = \lambda_5||\bm{x}_p - \hat{\bm{x}}_p||^2_2, 
\end{equation}
where $\lambda_5 > 0$ is a hyperparameter that controls the importance of this loss in the overall optimization.  

Finally, the overall training objective of the proposed approach is
\begin{equation} \psi^* = \arg\min_{\psi} \mathbb{E}_{(\bm{x}_h, \bm{x}_p) \sim \mathcal{X}_{\text{INN}}} \left[\mathcal{L}_{\text{forward}}+\mathcal{L}_{\text{backward}}+\mathcal{L}_{\text{privacy}} \right], 
\end{equation}
where we assume that the training dataset $\mathcal{X}_{\text{INN}}$ consists of a small subset of representative samples. We summarize the whole training and procedure in \autoref{algo:INN}. We note that this training process can be conducted offline at the sender or receiver, and does not require any prior knowledge about Eve's channel conditions, while only the semantic encoder and decoder are used. Once the training is completed, Alice and Bob can share the signal steganography module and realize secure SemCom.

\begin{figure*}[!t]
    \centering
    \begin{subfigure}[t]{0.30\textwidth}
        \centering
        \includegraphics[width=\textwidth]{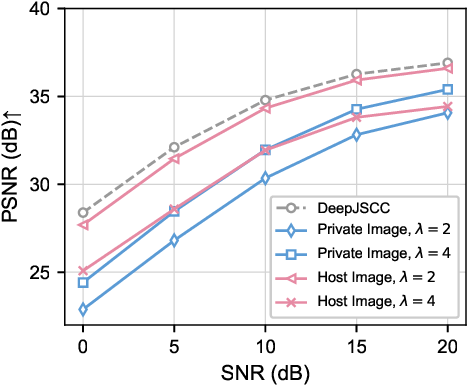}
        \caption{PSNR}
    \end{subfigure}
    \hfill
    \begin{subfigure}[t]{0.30\textwidth}
        \centering
        \includegraphics[width=\textwidth]{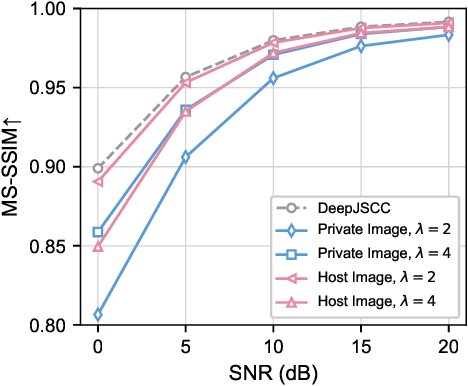}
        \caption{MS-SSIM}
    \end{subfigure}
        \hfill
    \begin{subfigure}[t]{0.30\textwidth}
        \centering
        \includegraphics[width=\textwidth]{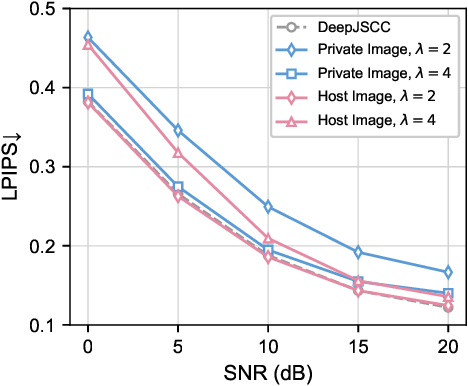}
        \caption{LPIPS}
    \end{subfigure}
    \caption{Performance of the proposed semantic covert communication under AWGN channels, where SNR varies from 0 dB to 20 dB.}
    \label{fig:AWGN_Bob_covert}
\end{figure*}
\begin{figure*}[!t]
    \centering
    \begin{subfigure}[t]{0.30\textwidth}
        \centering
        \includegraphics[width=\textwidth]{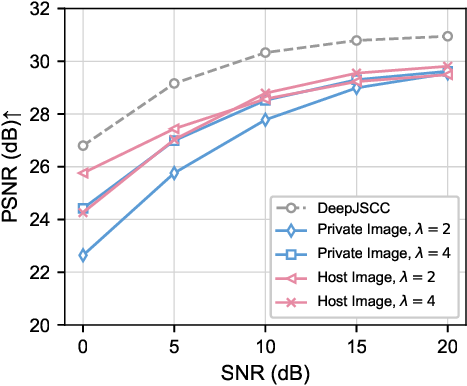}
        \caption{PSNR}
    \end{subfigure}
    \hfill
    \begin{subfigure}[t]{0.30\textwidth}
        \centering
        \includegraphics[width=\textwidth]{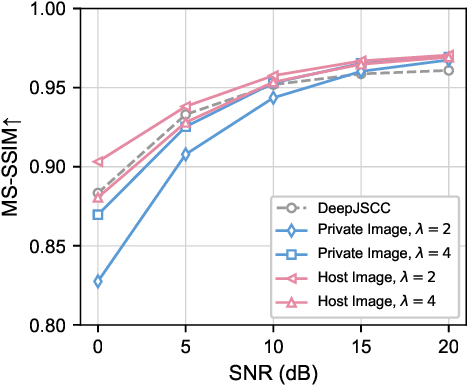}
        \caption{MS-SSIM}
    \end{subfigure}
        \hfill
    \begin{subfigure}[t]{0.294\textwidth}
        \centering
        \includegraphics[width=\textwidth]{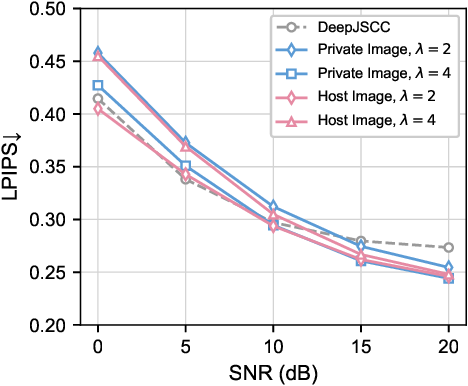}
        \caption{LPIPS}
        \label{fig:sub3}
    \end{subfigure}
    \caption{Performance of the proposed semantic covert communication under Rayleigh channels, where SNR varies from 0 dB to 20 dB.}
    \label{fig:rayleigh_Bob_covert}
\end{figure*}

\section{Simulations}
\label{Sec:simulations}
In this section, we conduct comprehensive simulations to evaluate the effectiveness of the proposed eavesdropping strategies and security mechanisms. Specifically, we first analyze the performance of the eavesdropping techniques to assess their ability to intercept semantic information. Then, we evaluate the proposed semantic covert communication scheme in countering these eavesdropping threats.  

\subsection{General Setup}
In simulations, we adopt the DeepJSCC architecture as our semantic encoder and decoder, with BCR set to 1/12. We consider both AWGN (i.e., $\bm{h}_b=\bm{1}$) and Rayleigh fading channels, which follow a complex Gaussian distribution $\mathcal{CN}(0, \bm{I})$. The SNR varies from 0 dB to 20 dB.  We evaluate the proposed eavesdropping and semantic cover communication under the CeleBAMask-HQ\cite{CelebAMask-HQ}, which consists of 30000 facial samples in total. We use 28000 samples to train DeepJSCC on SNRs uniformly sampled from 0 dB to 20 dB, and the rest 2000 samples for testing.  

To comprehensively evaluate the proposed methods, we use several metrics, including PSNR, multi-scale structural similarity index (MS-SSIM), and learned perceptual image patch similarity (LPIPS)\cite{LPIPS} from the perspectives of pixel-wise differences, structural details differences, and high-level perceptual quality. Moreover, according to \cite{weixuan_Differential_Privacy,fppsr}, we also evaluate the ability of Eve to recover facial privacy information from intercepted signals via a face recognition system named ArcFace\cite{ArcFace}. Specifically, we define the facial privacy eavesdropping success rate (FPESR) as the proportion of the recovered images by Eve that are correctly identified as belonging to the same person as the transmitted one. It is noted that a larger FPESR indicates a higher risk of privacy leakage, while a lower one suggests stronger privacy protection in SemCom systems.

\subsection{Evaluation of the Proposed Eavesdropping}
 We first evaluate the effectiveness of the proposed eavesdropping strategies, including glass-box, closed-box, and their improved versions, referred to as GenAI+glass-box and GenAI+closed-box, respectively. We use the SemanticStylGAN\cite{SemanticStyleGAN} as the generator of the GenAI+glass-box and GenAI+closed-box methods. All methods are tested under additive white Gaussian noise (AWGN) channels with SNR of 0 dB to 20 dB. Moreover, for the glass-box and GenAI+glass-box methods, we adopt the Adam with a learning rate of $1\times 10^{-3}$ to iteratively optimize the initial image. For closed-box and GenAI+closed-box, we set the number of APIs to 100 and also use the Adam optimizer with a learning rate of $1\times 10^{-3}$ to optimize the inversion network. For reference, we also provide the results in the scenario that Eve has the semantic decoder to directly recover the private image.

\autoref{fig:AWGN_Eve} and \autoref{fig:Rayleigh_Eve} show the PSNR, MS-SSIM, and LPIPS results of different eavesdropping strategies on AWGN channels and Rayleigh channels, respectively. From these figures, we can observe that the performance of the proposed eavesdropping strategies improved with an increased SNR. Specifically, the glass-box achieves the worst eavesdropping performance, because it has the least information about the original distribution of the transmitted images. However, it still achieves a PSNR of about 15dB and MS-SSIM of 0.6 on AWGN channel when SNR is 0dB, and a PSNR of about 10dB and MS-SSIM of 0.4 on the Rayleigh channel. Moreover, with the introduction of the GenAI model, the eavesdropping performance is significantly improved and achieves the best LPIPS performance among all methods. For closed-box eavesdropping, although it cannot access the architecture of the semantic encoder, it can still achieve better performance than glass-box eavesdropping, and GenAI can also improve the performance of closed-box eavesdropping. These results can demonstrate the effectiveness of the proposed eavesdropping strategies.

\autoref{fig:fpesr} shows the FPESR of different eavesdropping strategies on AWGN channels and Rayleigh channels with SNR of 5dB. We can find that despite not achieving high image quality metrics, the proposed glass-box achieves above 80\% on both types of channels, which indicates that the glass-box eavesdropping can still risk the privacy of the transmitted images. Moreover, the closed-box, GenAI+glass-box, and GenAI+closed-box eavesdropping improve the FPESR to 100\%  which further demonstrates the effectiveness of the proposed eavesdropping strategies.

To further illustrate the effectiveness of our proposed approaches, we provide visual examples of reconstructed images using different eavesdropping strategies in \autoref{fig:vis_eve}. From this figure, we can observe that although the reconstructed image by the glass-box method exhibits noticeable blurriness, noise artifacts, and distortions, it still preserves distinguishable facial features and identity characteristics of the original subjects. This explains why the glass-box eavesdropping achieves high FPESR despite its low PSNR and MS-SSIM scores. Moreover, the closed-box method produces images with less noise compared to the glass-box approach, and with the incorporation of GenAI models, both glass-box and closed-box methods show significant improvements in reconstruction quality, particularly in preserving facial details.

\subsection{Evaluation of Semantic Covert Communication}
Next, we evaluate the performance of the proposed semantic covert communication scheme. Specifically, the number of invertible blocks in the INN-based signal steganography module is set to 8. To simulate the real scenario, we assume that only 1000 samples are used in the training process of the signal steganography module. We train the signal steganography module using the Adam optimizer with a learning rate of $3\times 10^{-4}$ and a batch size of 128. The values of hyperparameters $\lambda_1$, $\lambda_2$, $\lambda_3$, $\lambda_4$, and $\lambda_5$ are set to 1, 2, 2, 1, and 1, respectively, unless otherwise specified.

In \autoref{fig:AWGN_Bob_covert} and \autoref{fig:rayleigh_Bob_covert}, we first show the impact of the proposed semantic covert communication scheme on the legitimate receiver, Bob, under AWGN and Rayleigh channels, respectively, where $\lambda_3$ is set to 2 and 3. For reference, we include the performance of DeepJSCC without any secure mechanisms. From these figures, we can observe that the proposed approach maintains a comparable reconstruction quality of the private image compared to the scenario without any secure mechanisms. Specifically, the reconstruction quality of the host image is very close to that of DeepJSCC under AWGN channels, where there is only less than 1dB reduction in PSNR and 0.05 reduction in MS-SSIM when $\lambda_3=2$. Besides, the reconstruction quality of the private image is also well preserved and achieves a performance similar to that of DeepJSCC in terms of LPIPS. Moreover, in Rayleigh channels, although the PSNR of the host image and the private image gets reduced, the MS-SSIM and LPIPS scores are still comparable with DeepJSCC and even outperform DeepJSCC in terms of LPIPS when SNR is larger than 10dB. This is because the invertible blocks in the INN-based signal steganography module can effectively mitigate the errors introduced by channel fading and channel noise. These results demonstrate that the proposed semantic covert communication does not deteriorate the reconstruction quality of the host and private images at the legitimate receiver side.
\begin{table*}[!t]
    \centering
    \caption{Performance of the proposed semantic covert communication under different eavesdropping strategies under AWGN channels with SNR of the eavesdropping channel set to 20 dB.}
    \label{tab:Eve_covert}
    \renewcommand{\arraystretch}{1.2} 
    \setlength{\tabcolsep}{7pt} 
    \begin{tabularx}{0.88\linewidth}{lccc|ccc|c}
        \toprule
        \multirow{2}{*}{\textbf{Method}} & \multicolumn{3}{c|}{\textbf{Similarity with Host Images}} & \multicolumn{3}{c|}{\textbf{Similarity with Private Images}} & \multirow{2}{*}{\textbf{FPESR }(↓)} \\
        \cline{2-4} \cline{5-7}
        & \textbf{PSNR} (↑) & \textbf{MS-SSIM} (↑) &\textbf{ LPIPS} (↑) &\textbf{ PSNR} (↓) & \textbf{MS-SSIM} (↓) & \textbf{LPIPS} (↓) & \\
         \midrule
        DeepJSCC & 32.85dB & 0.9856 & 0.1638 & 9.07dB & 0.3269 & 0.6594 & 0.0 \\
        \midrule
        Glass-box & 15.87dB & 0.6992 & 0.7311 & 9.08dB & 0.2277 & 0.7973 & 0.0 \\
        Closed-box & 27.50dB & 0.9595 & 0.4129 & 9.05dB & 0.3328 & 0.6940 & 0.0 \\
        GenAI+Glass-box & 24.16dB & 0.8988 & 0.4051 & 9.34dB & 0.3353 & 0.6881 & 0.0 \\
        GenAI+Closed-box & 24.60dB & 0.9075 & 0.5053 & 9.18dB & 0.3227 & 0.6878 & 0.0 \\
        \bottomrule
    \end{tabularx}
\end{table*}
\begin{figure*}
 \centering
        \includegraphics[width=0.88\textwidth]{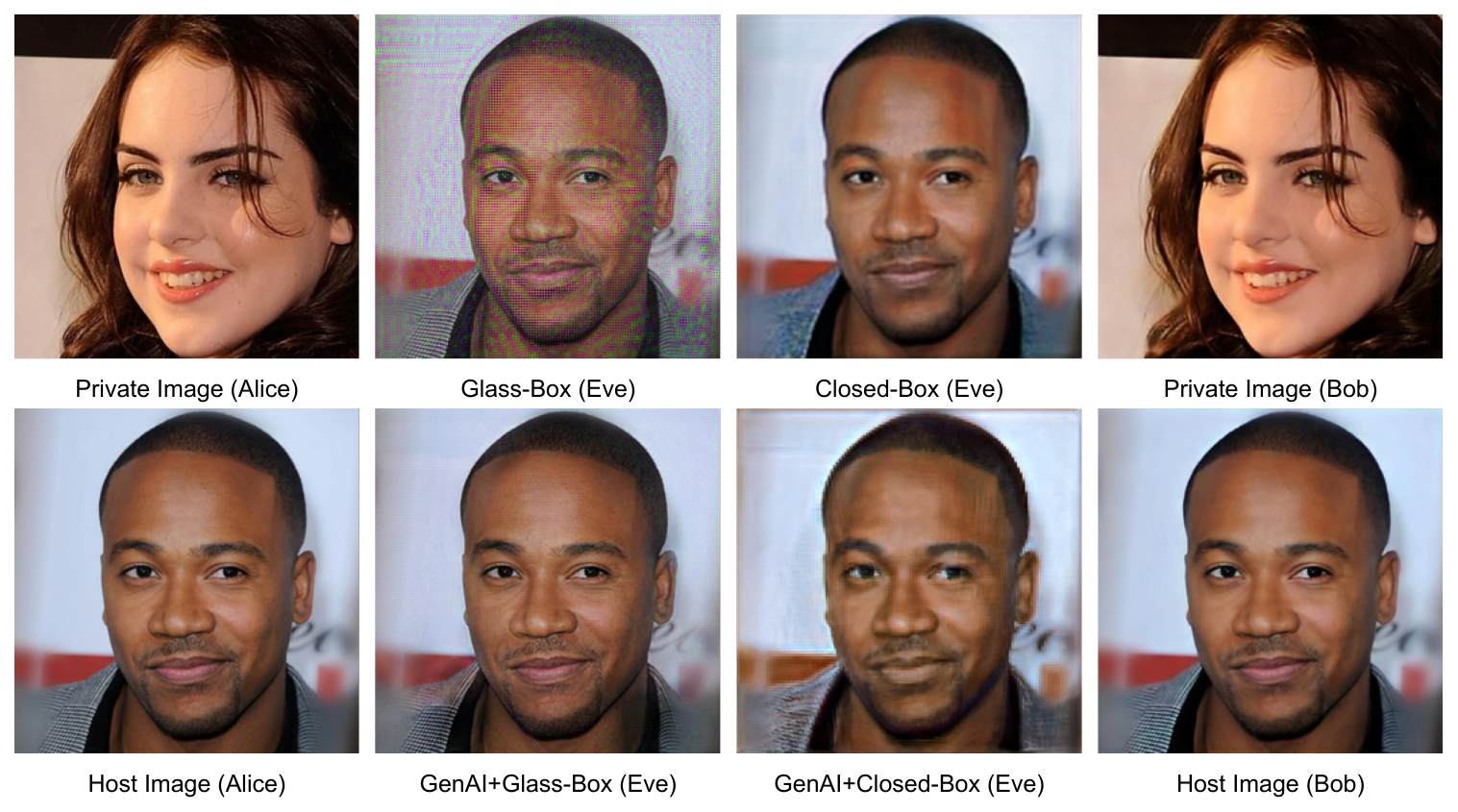}
        \caption{Visualization of the reconstructed images at Eve and Bob with the proposed semantic covert communication scheme under AWGN channels with SNR of 20dB. }
        \label{fig:Vis_Covert}
\end{figure*}

As shown in \autoref{tab:Eve_covert}, we evaluate the security of the proposed semantic covert communication against different eavesdropping strategies. For closed-box and GenAI+closed-box eavesdropping, we assume that Eve can still observe the output of the INN-based signal steganography module and train the inverse network to recover the original image. Moreover, we also provide the performance of the DeepJSCC, where Eve can directly recover the private image using the semantic decoder for reference. From the results in this table, we can observe that Eve can only reconstruct the host image and achieve high enough PSNR, MS-SSIM, and LPIPS scores, while it cannot obtain the private image. Moreover, the FPESR of all eavesdropping strategies is reduced to 0, indicating that the proposed semantic covert communication scheme effectively protects the privacy of the transmitted images. These results demonstrate the effectiveness of the proposed semantic covert communication scheme in enhancing the security of SemCom systems against intelligent eavesdroppers.

\autoref{fig:Vis_Covert} illustrates the reconstructed images by different eavesdropping strategies with the proposed semantic covert communication scheme under AWGN channels with an SNR of 20dB. From this figure, we can find that the images reconstructed by Eve using any eavesdropping strategies are significantly different from the private images, but are very similar to the host images without over-distortion. This can effectively protect the privacy of the transmitted images. Moreover, Bob can reconstruct both the host and private images with high quality, which further demonstrates the effectiveness of the proposed semantic covert communication scheme.

\section{Conclusion}
\label{Sec:conclusion}
In this paper, we have explored the security challenges in SemCom systems, particularly in the presence of intelligent eavesdroppers. We have proposed novel eavesdropping strategies, including both glass-box and closed-box attacks, and demonstrated their effectiveness in intercepting semantic information. To counter these threats, we have introduced a semantic covert communication framework using signal steganography. Our simulation results have shown that these defense strategies significantly enhance the security of SemCom systems, effectively mitigating the risks posed by intelligent eavesdroppers. 
\bibliographystyle{IEEEtran}
\bibliography{IEEEabrv, ref}

\end{document}